# Photoresponse of $La_{1.85}Sr_{0.15}CuO_4$ nanostrip

**Hiroyuki Shibata[1], Naoto Kirigane[1], Kentaro Fukao[1], Daisuke Sakai[1], Shinichi Karimoto[2], and Hideki Yamamoto[2]**

[1] Kitami Institute of Technology, 165 Koen-cho, Kitami, Hokkaido 090-8507, Japan

[2] NTT Basic Research Laboratories, 3-1, Wakamiya, Morinosato, Atsugi-shi, Kanagawa 243-0198, Japan

E-mail: shibathr@mail.kitami-it.ac.jp

Abstract

We report the fabrication and photoresponse of 5nm thick $La_{1.85}Sr_{0.15}CuO_4$ nanostrip with a width of 100nm. The I-V characteristics of the nanostrip show a hysteresis and a sharp voltage jump at $I_c$. The $J_c$(3K) of the nanostrip is $2.3 \times 10^7$ A/cm$^2$. The nanostrip exhibits photoresponse signals when illuminated by a pulse laser at 1560 nm wavelength with a bias current just below $I_c$. The height of the signal reduces as the optical intensity decreases and disappears below -10 dBm. The signal also decreases as the temperature increases, but it exists up to 30 K. These results suggest the possibility of $La_{1.85}Sr_{0.15}CuO_4$ nanostrip as single-photon detector (SSPD, SNSPD) working at high temperature by further reducing the cross-section of the strip.

## 1. Introduction

High-performance single-photon detector in the infrared region is required in many fields such as quantum information, quantum optics, space-to-ground communications, and bio sensing [1]. For this purpose, superconducting single-photon detector (SSPDs, SNSPDs) has been attracted much attention in the last decade [1,2]. Now the SSPDs with the detection efficiency close to unity (93%) have been reported by several groups [3,4]. It is also noted that the dark count rate of SSPD has been reached to a theoretical limit of background blackbody radiation [5,6]. By introducing cold bandpass filters with 100 GHz bandwidth, the dark count rate of SSPD decreases to $10^{-4}$ cps [6]. The SSPD with ultralow dark count rate is particularly important for the long distance quantum key distribution (QKD) over 300 km [7].

Although the performance of SSPDs overcomes to the other single-photon detectors such as InGaAs avalanche photodiode, a drawback is a low operating temperature. The operating temperature is below 3 K for NbN based SSPD, and below 1 K for WSi based SSPD, depending on the $T_c$ of the materials [1-6]. To increase the operating temperature, we have to use higher $T_c$ materials for SSPDs. Recently, we have developed the SSPDs based on $MgB_2$ ($T_c$ = 39 K) [8], and it has been demonstrated that the $MgB_2$-SSPD exhibits the capacity of detecting single photons up to 10 K [9,10]. To further increase the operating temperature, it is inevitable to use cuprate superconductors for SSPDs.

One of the key technological issues for the realization of cuprate based SSPDs is the ultrathin film growth. We need high quality ultrathin films for SSPDs; with the thickness of about 5nm, a sharp superconducting transition, and a smooth surface. There have been several reports towards the realization of SSPDs using $YBa_2Cu_3O_{7-\delta}$ films with $T_c$ about 90 K [11-15]. However, as the thickness decreases to 5 nm, the quality of the films strongly deteriorates [14]. Arpaia et al. fabricated the $YBa_2Cu_3O_{7-\delta}$ parallel type meander structure down to 100 nm wide ×50 nm thick and observed the photoresponse of the meanders [13]. Amari et al. fabricated the $YBa_2Cu_3O_{7-\delta}$ meanders with 100 nm wide × 30 nm thick by ion irradiation [15]. It is necessary to further reduce the thickness of the film for realizing the cuprate based SSPDs.

Among the cuprates, there is a $La_{1.85}Sr_{0.15}CuO_4$ with $T_c$ of 37 K. It has been reported that it is possible to grow high quality epitaxial films of $La_{1.85}Sr_{0.15}CuO_4$ with the thickness of 5 nm on a $LaSrAlO_4$ substrate [16-18]. So, if we use $La_{1.85}Sr_{0.15}CuO_4$ instead of $YBa_2Cu_3O_{7-\delta}$, we can solve the problem of ultrathin film growth for SSPDs.

Here, we report the fabrication of $La_{1.85}Sr_{0.15}CuO_4$ nanostrip with the size of 100 nm wide × 10 μm long × 5 nm thick. From the transport measurement, it is revealed that the strip has good superconducting characteristics. The strip shows a photoresponse up to 30 K, suggesting the possible single-photon detection at high temperature in the future.

## 2. Fabrication

Ultrathin films of $La_{1.85}Sr_{0.15}CuO_4$ were synthesized in a custom-designed ultrahigh-vacuum chamber (base pressure $<10^{-7}$ Pa) from metal sources by means of reactive co-evaporation [16,17,19].

For oxidation of the film, RF activated atomic oxygen (input power of 300 W and oxygen flow rate of 1 sccm) was used. During the growth, the evaporation beam flux of each element was controlled by electron impact emission spectroscopy (EIES) and the film was monitored by reflection high-energy electron diffraction (RHEED). The growth rate was 9 nm/min and the growth temperature is 670 °C. The 5 nm thick thin films with (001) orientation were epitaxially grown on (001) LaSrAlO$_4$ substrates. Details of our growth technique have been published elsewhere [16,17,19].

Figure 1 shows the resistivity vs temperature ($\rho$-T) curves of La$_{1.85}$Sr$_{0.15}$CuO$_4$ thin film with the thickness of 5 nm. As shown in the inset of Fig. 1, T$_c$ (R = 0) is 41.6 K, which is higher than the value (37 K) of bulk specimens. This is due to the compressive strain effect caused by the lattice mismatch between film and substrate [16,17]. The in-plane lattice parameter of LaSrAlO$_4$ is 0.3756 nm, which is about 5% shorter than the 0.3777 nm of La$_{1.85}$Sr$_{0.15}$CuO$_4$. Now, we can use very high quality ultrathin films with T$_c$ above 40 K for the development of SSPDs.

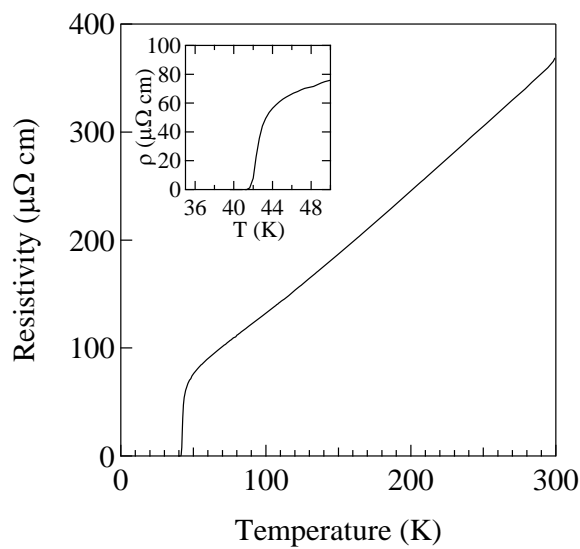

**Figure 1.** Temperature dependence of the resistivity of 5 nm thick La$_{1.85}$Sr$_{0.15}$CuO$_4$ thin films. Insets: superconducting transition.

For the nanofabrication of cuprates, Au passivation is commonly used to protect the damage. However, it is impossible to use conductive materials as a passivation layer in the case of SSPDs. Here, we use AlN for the passivation of La$_{1.85}$Sr$_{0.15}$CuO$_4$ films. The AlN film has been known to be useful as a passivation layer of MgB$_2$-SSPDs [20]. The single line pattern with 10 μm long and 100 nm wide was fabricated on silicon-based negative resist (SNR) by e-beam lithography. Figure 2 shows scanning electron microscopy (SEM) images of the pattern. Then the pattern was transferred to the La$_{1.85}$Sr$_{0.15}$CuO$_4$ film using standard Ar ion milling. In this way, we can fabricate La$_{1.85}$Sr$_{0.15}$CuO$_4$ strip with the size of 100 nm wide × 10 μm long × 5 nm thick.

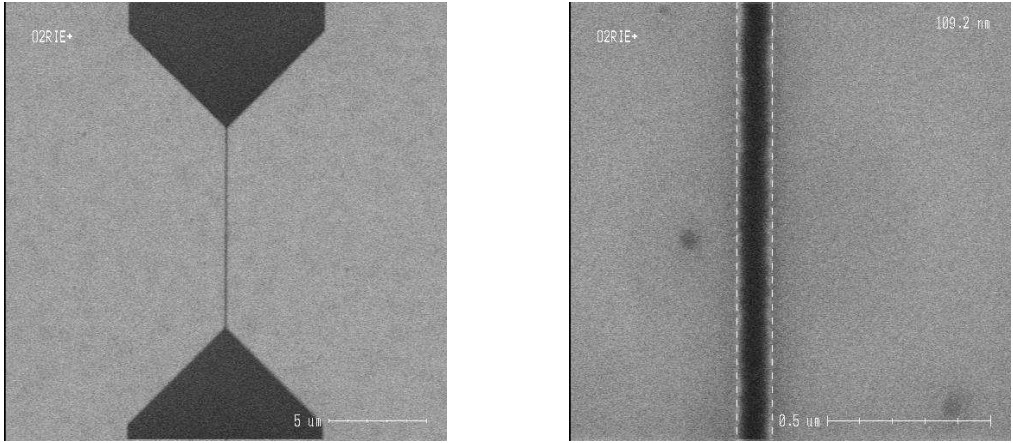

**Figure 2. (a)** SEM image of $La_{1.85}Sr_{0.15}CuO_4$ strip with the size of 100 nm wide × 10 μm long × 5 nm thick before milling and **(b)** an enlarged view.

## 3. I-V characteristics

Figure 3 shows the current-voltage (I-V) characteristics of the strip at various temperatures using two terminal methods. At 3 K, it shows a voltage jump at a critical current ($I_c$) and a small hysteresis behavior (inset). These behaviors are different to flux-flow type behaviors which are frequently observed in inhomogeneous cuprate nanostructures, and suggest the high uniformity of the strip. The $I_c$ is 115 μA, which corresponds to the critical current density ($J_c$) of $2.3 \times 10^7$ A/cm$^2$. The large $J_c$ value also indicates the high quality of the strip. The $I_c$ structure can be observed clearly up to 35 K, and the slope of I-V curve strongly decreases at 45 K corresponding to the normal resistance. These features clearly show that it is possible to fabricate the high quality $La_{1.85}Sr_{0.15}CuO_4$ nanostrip with the size of 100 nm wide × 10 μm long × 5 nm thick without degradation.

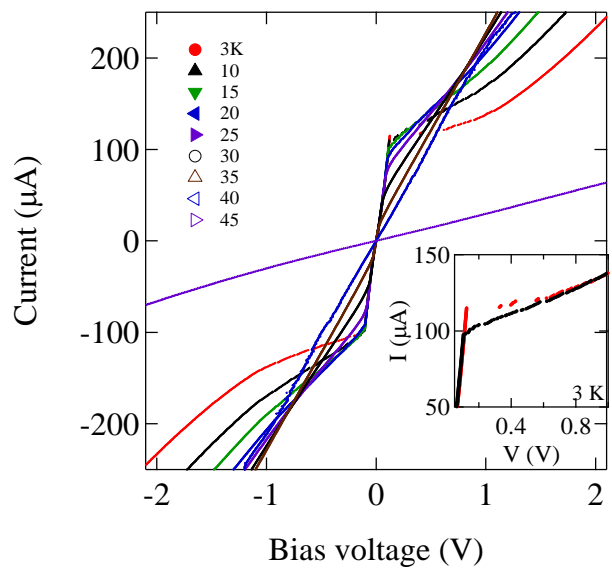

**Figure 3.** I-V characteristics of $La_{1.85}Sr_{0.15}CuO_4$ strip with the size of 100 nm wide × 10 μm long × 5 nm thick at various temperatures. Insets: Hysteresis behavior at 3 K.

## 4. Photoresponse

Electro-optical characterization was performed in a Gifford-McMahon cryocooler between 3 K and 45 K. A bias current was supplied to the strip through a bias-T without shunt resistor, and the signal was amplified by an rf amplifier with a total gain of 30 dB and fed to a 20 GHz bandwidth single-shot oscilloscope [21]. The strip was illuminated with a femtosecond 1560 nm fiber laser with a repetition rate of 100 MHz. The optical spot size was about 10 μm in diameter.

No signals are observed at zero bias current ($I_{bias}$ = 0), but the signal appears as the bias current increases, and the height of the signal becomes maximum at the bias current just below $I_c$. Figure 4 shows the optical power dependence of the transient photoresponse of $La_{1.85}Sr_{0.15}CuO_4$ strip at 3 K. The $I_{bias}$ is 111 μA, just below $I_c$. The height of the signal decreases as the optical intensity decreases, and disappears below -10 dBm. The estimated incident number of photons on the nanostrip is $9.9 \times 10^4$ photons/pulse at -10dBm, and it is far from single-photon detection regime.

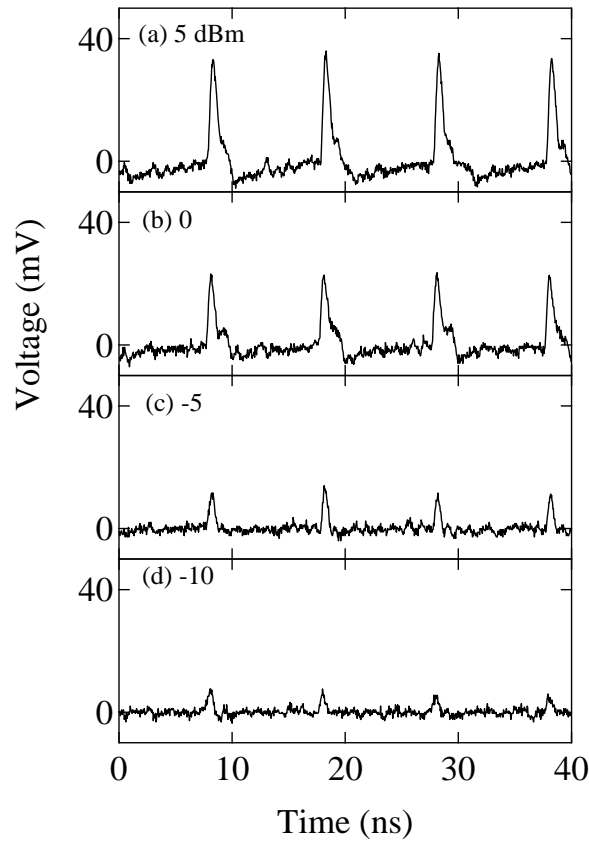

**Figure 4.** Transient photoresponse of $La_{1.85}Sr_{0.15}CuO_4$ strip with the size of 100 nm wide × 10 μm long × 5 nm thick with various laser pulse power. The strip is illuminated with a 1.56 μm wavelength femtosecond laser pulse at a 100 MHz repetition rate and measured at 3 K with $I_{bias}$ = 111 μA.

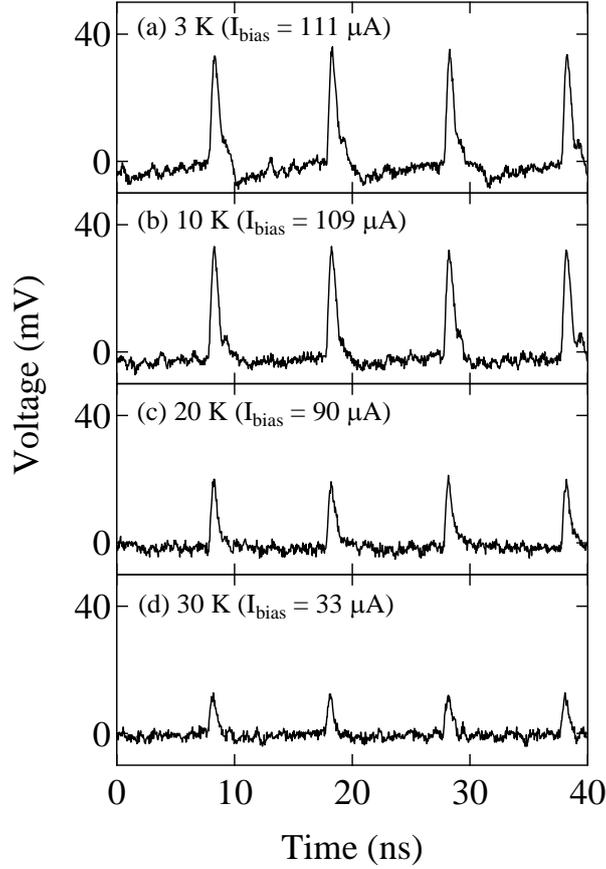

**Figure 5.** Transient photoresponse of $La_{1.85}Sr_{0.15}CuO_4$ strip with the size of 100 nm wide × 10 μm long × 5 nm thick with various temperatures. The strip is illuminated with a 1.56 μm wavelength femtosecond laser pulse at a 100 MHz repetition rate and $I_{bias}$'s are adjusted just below $I_c$.

Figure 5 shows the temperature dependence of the transient photoresponse of $La_{1.85}Sr_{0.15}CuO_4$ strip. Here, the bias currents are adjusted just below $I_c$ at each temperature. As the temperature increases, the height of the signal decreases, but it can be observed up to 30 K. The decrease of the height is simply due to the decreases of $I_{bias}$.

Although the reason of these photoresponse is not clear, it seems to be a bolometric response of the nanostrip under the bias current. We notice that the small signals remain even when $I_{bias}$ exceeds $I_c$, which may be due to the large temperature coefficient of the resistivity in the normal state. It is also noted that the same response have been also observed in the $MgB_2$ nanostrip with the width of 300 nm [22]. The response of $MgB_2$ nanostrip changes to the multi-photon detection regime with the width of 200 nm and changes to the single-photon detection regime with the width of 100 nm [8,22]. The bolometric response seems a common feature for superconducting nanostrip with the large cross-section.

Finally, we discuss the possibility towards single-photon detection of $La_{1.85}Sr_{0.15}CuO_4$ strip. Due to the high $J_c$ of $La_{1.85}Sr_{0.15}CuO_4$, the $I_c$ of the present strip is about 5 times larger than the standard SSPD using other material (about 20 μA). So, we may have to reduce the cross section of our nanostrip about 5 times for single photon detection. The homogeneous nanostrip with 20 nm wide is required when the 5 nm thick $La_{1.85}Sr_{0.15}CuO_4$ film is used. The requirement is quite challenging and needs further technological progresses. At the present stage, cuprate based SSPDs may be useful to detect high energy particles, such as single photon in the X ray region and single biomolecular ion [23,24]. Actually, in the case of $MgB_2$, single biomolecular ion detection has been reported with 100 % detection efficiency up to 13 K [23].

## 5. Conclusion

We grow 5 nm thick $La_{1.85}Sr_{0.15}CuO_4$ film using MBE and fabricate a strip with the size of 100 nm wide × 5 nm thick with $J_c(3K) = 2.3 \times 10^7$ A/cm$^2$. The I-V characteristics show a hysteresis and a sharp voltage jump at $I_c$. We observe the photoresponse of the strip under bias current even at 30 K, which is due to the bolometric response. We discuss the possible single-photon detection using $La_{1.85}Sr_{0.15}CuO_4$ nanostrip at high working temperature.

## Acknowledgements

The authors would like to thank A. Tsukada, M. Naito, and H. Sato for many useful discussions about ultrathin film growth of $La_{1.85}Sr_{0.15}CuO_4$.